\documentstyle[12pt]{article}
\def\1{\mbox{I\hspace{-.15em}1}}

\def\b{\begin{equation}}
\def\e{\end{equation}}
\def\bee{\begin{enumerate}}
\def\eee{\end{enumerate}}

\title{Torsion of Space-time in $ f(R) $ gravity}
\author{M. Mohsenzadeh$^{1}$\thanks{e-mail:
mohsenzadeh@qom-iau.ac.ir} and E. Yusofi$^{2}$\thanks{e-mail:
e.yusofi@iauamol.ac.ir}}
\begin{document}

\maketitle {\it \centerline{\it $^{1}$ Department of Physics, Qom
Branch, Islamic Azad University, Qom, Iran }  \centerline{$^2$
Department of Physics, Ayatollah Amoli Branch, Islamic Azad University, Iran}
\centerline{\it P.O.BOX 678, Amol, Mazandaran}}

\begin{abstract}
In this paper, we first review some aspects of the $ f(R) $
gravity and then the concept of torsion of space-time due to
metric-affine formalism in $ f(R) $ gravity is studied. Within
this formalism in which the matter action is supposed to dependent
on the connection, we achieve to interesting cases including
non-zero torsion tensor. Then with the physical interpretation of
torsion of space-time in high energy limit, the modified
expression of Mach's principle in a very strong gravitational
region is obtained.

\end{abstract}


\section{Introduction and Motivation}
General Relativity (GR) is widely accepted as a fundamental theory
to describe the geometric properties of space-time. In a
homogeneous and isotropic space-time the Einstein field equations
give rise to the Friedmann equations that describe the evolution
of the universe. In fact, the standard big-bang cosmology based on
radiation and matter dominated epochs can be well described within
the framework of GR.

However, the rapid development of observational cosmology which
started from 1990s shows that the universe has undergone two
phases of cosmic acceleration. The first one is called inflation,
which is believed to have occurred prior to the radiation
domination. This phase is required not only to solve the flatness
and horizon problems plagued in big-bang cosmology, but also to
explain a nearly flat spectrum of temperature anisotropies
observed in Cosmic Microwave Background (CMB). The second
accelerating phase has started after the matter domination
\cite{c1}. The unknown component giving rise to this late-time
cosmic acceleration is called dark energy \cite{c2}. The existence
of dark energy has been confirmed by a number of observations such
as supernovae Ia (SN Ia), large scale structure (LSS), baryon
acoustic oscillations (BAO), and CMB.

These two phases of cosmic acceleration cannot be explained by the presence of standard matter
whose equation of state $ \omega=\frac{P}{\rho} $ satisfies the condition $ \omega\geq 0 $ (here $P $ and $ \rho $ are the pressure
and the energy density of matter, respectively). In fact, we further require some component of
negative pressure, with $ \omega < -\frac{1}{3} $, to realize the acceleration of the universe. The cosmological
constant $ \Lambda $ is the simplest candidate of dark energy, which corresponds to $ \omega=-1 $. However, if the
cosmological constant originates from a vacuum energy of particle physics, its energy scale is too
large to be compatible with the dark energy density. Hence we need to find some mechanism to
obtain a small value of $ \Lambda $ consistent with observations. Since the accelerated expansion in the very
early universe needs to end to connect to the radiation-dominated universe, the pure cosmological
constant is not responsible for inflation. A scalar field $ \phi $ with a slowly varying potential can be a
candidate for inflation as well as for dark energy.

Although many scalar-field potentials for inflation have been constructed in the framework
of string theory and supergravity, the CMB observations still do not show particular evidence
to favor one of such models. This situation is also similar in the context of dark energy-there
is a degeneracy as for the potential of the scalar field (quintessence) due to the observational degeneracy to the dark energy equation of state around
$ \omega=-1 $. Moreover it is generally difficult to construct viable quintessence potentials motivated
from particle physics because the field mass responsible for cosmic acceleration today is very small
($ m_{\phi}\simeq 10^{-33}eV $).

While scalar-field models of inflation and dark energy correspond to a modification of the
energy-momentum tensor in Einstein equations, there is another approach to explain the acceleration
of the universe. This corresponds to the modified gravity in which the gravitational theory is
modified compared to GR. The Lagrangian density for GR is given by $ f(R) = R-2\Lambda $, where $ R $ is
the Ricci scalar and $ \Lambda $ is the cosmological constant (corresponding to the equation of state $ \omega=-1 $).
The presence of $ \Lambda $ gives rise to an exponential expansion of the universe, but we cannot use it for
inflation because the inflationary period needs to connect to the radiation era. It is possible to use
the cosmological constant for dark energy since the acceleration today does not need to end. However,
if the cosmological constant originates from a vacuum energy of particle physics, its energy
density would be enormously larger than the today's dark energy density. While the $ \Lambda $-Cold Dark
Matter ($\Lambda $CDM) model ($ f(R) = R-2\Lambda $) fits a number of observational data well, there
is also a possibility for the time-varying equation of state of dark energy.

One of the simplest modifications to GR is the $ f (R) $ theories
of gravity in which the Lagrangian density is supposed to be an
arbitrary function of $ R $ \cite{c3,c4}. The $ f(R) $ theories of
gravity come about by a straightforward generalization of the
Lagrangian in the Einstein-Hilbert action, \b
S_{EH}=\frac{1}{2k}\int d^{4}x\sqrt{-g}R, \e where $ k\equiv 8\pi
G $, $ G $ is the gravitational constant, $ g $ is the determinant
of the metric $ g_{\mu\nu}$ and $ R $ is the Ricci scalar ($
c=\hbar=1 $ ), to become a general function of $ R $, i.e., \b
S=\frac{1}{2k}\int d^{4}x\sqrt{-g}f(R).\e

As can be found in many textbooks - see, for example\cite{c5,c6}-
there are actually two variational principles that one can apply
to the Einstein-Hilbert action in order to derive Einstein's
equations: the standard metric variation and a less standard
variation dubbed Palatini variation [even though it was Einstein
and not Palatini who introduced it \cite{c7}]. In the former the
metric is assumed to be independent variable but in the latter the
metric and the connection are assumed to be independent variables
and one varies the action with respect to both of them, under the
important assumption that the matter action does not depend on the
connection. The choice of the variational principle is usually
referred to as a formalism, so one can use the terms metric (or
second order) formalism and Palatini (or first order) formalism.
Therefore, it is intuitive that there will be two version of
$f(R)$ gravity, according to which variational principle or
formalism is used. Indeed this is the case: $ f(R) $ gravity in
the metric formalism is called metric $ f(R) $ gravity and $ f(R)
$ gravity in the Palatini formalism is called Palatini $ f(R) $
gravity \cite{c8}.

Finally, there is actually even a third version of $ f(R) $
gravity: metric-affine $ f(R) $ gravity \cite{c9,c10}. This comes
about if one uses the Palatini variation but abandons the
assumption that the matter action is independent of the
connection. Clearly, metric-affine $ f(R) $ gravity is the most
general of these theories and reduces to metric or Palatini $f(R)$
gravity if further assumptions are made.

In this paper, we first study the formalism of modified gravity, i.e. $ f (R) $ gravity.
Then we are going to express a modified form of Mach's principle with a closer look at the
concepts of curvature and torsion. Therefore, in section 2, we study the metric-affine formalism of modified
gravity in the detailed review. We take a closer look in the concept of torsion of space-time and correct expression of
Mach's principle in section 3. We bring a summary of the results in the final section.

\section{Metric-Affine formalism of $ f(R) $ gravity}

As we pointed out, the Palatini formalism of $ f(R) $ gravity
relies on the crucial assumption that the matter action does not
depend on the independent connection. This assumption relegates
this connection to the role of some sort of auxiliary field and
the connection carrying the usual geometrical meaning - parallel
transport and definition of the covariant derivative - remains the
Levi-Civita connection of the metric \cite{c11}. But if we decided
to be faithful to the geometrical interpretation of the
independent connection $ \Gamma^{\lambda}_{\mu\nu} $, then this
would imply that we would define the covariant derivatives of the
matter fields with this connection and, therefore, we would have
\b S_{M}=S_{M}(g_{\mu\nu},\Gamma^{\lambda}_{\mu\nu},\psi), \e
where $ \psi $  collectively denotes the matter fields. The action
of this theory, dubbed metric-affine $ f(R) $ gravity \cite{c10},
would then be \b S_{ma}=\frac{1}{2k}\int
d^{4}x\sqrt{-g}f(\Re)+S_{M}(g_{\mu\nu},\Gamma^{\lambda}_{\mu\nu},\psi).\e
where $ f(\Re) $ is a general function of $ \Re $, and the Ricci
scalar $ \Re $ is constructed with the independent connection $
\Gamma^{\lambda}_{\mu\nu} $.

Before going further and deriving field equations from this action
certain issues need to be clarified. First, since now the matter
action depends on the connection, we should define a quantity
representing the variation of $ S_{M} $ with respect to the
connection, which mimics the definition of the stress-energy
tensor. We call this quantity the hyper-momentum and is defined as
\cite{c12} \b \Delta_{\lambda}^{\mu\nu}\equiv
-\frac{2}{\sqrt{-g}}\frac{\delta S_{M}}{\delta
\Gamma^{\lambda}_{\mu\nu}}.\e

Additionally, since the connection is now promoted to
the role of a completely independent field, it is interesting
to consider not placing any restrictions to it. Therefore,
besides dropping the assumption that the connection is
related to the metric, we will also drop the assumption
that the connection is symmetric. Also, it is useful to define
the Cartan torsion tensor \b {S_{\mu\nu}}^{\lambda}\equiv {\Gamma^{\lambda}}_{[\mu\nu]}, \e
which is the anti-symmetric part of the connection. $ [\mu\nu] $ denote anti-symmetrization over the indices $ \mu $ and
$ \nu $.

By allowing a non-vanishing Cartan torsion tensor we are allowing
the theory to naturally include torsion. Even though this brings
complications, it has been considered by some to be an advantage
for a gravity theory since some matter fields, such as Dirac
fields, can be coupled to gravity in a way which might be
considered more natural\cite{c13}: one might expect that at some
intermediate or high energy regime, the spin of particles might
interact with the geometry (in the same sense that macroscopic
angular momentum interacts with geometry) and torsion can
naturally arise. Theories with torsion have a long history,
probably starting with the Einstein-Cartan(-Sciama-Kibble)
theory\cite{c14,c15}. In this theory, as well as in other theories
with an independent connection, some part of the connection is
still related to the metric (e.g., the non-metricity is set to
zero). In our case, the connection is left completely
unconstrained and is to be determined by the field equations.
Metric-affine gravity with the linear version of the action (4)
was initially proposed in\cite{c12} and the generalization to $
f(\Re) $ actions was considered in \cite{c9,c10}.

The final form of the field equations is\cite{c11}:
\b f'(\Re)\Re_{(\mu\nu)}-\frac{1}{2}f(\Re)g_{\mu\nu}=kT_{\mu\nu},\e
\b \frac{1}{\sqrt{-g}}[-\bar{\nabla}_{\lambda}(\sqrt{-g}f'(\Re)g^{\mu\nu})+\bar{\nabla}_{\sigma}(\sqrt{-g}f'(\Re)g^{\mu\sigma}){\delta^{\nu}}_{\lambda}]
+2f'(\Re)g^{\mu\sigma}{S_{\sigma\lambda}}^{\nu}=k({\Delta_{\lambda}}^{\mu\nu}-\frac{2}{3}{\Delta_{\sigma}}^{\sigma[\nu}{\delta^{\mu]}}_{\lambda},\e
\b {S_{\mu\sigma}}^{\sigma}=0\e
where $ T_{\mu\nu} $ is stress-energy
tensor, prime denote the variation with respect to the metric, $ \bar{\nabla} $ denotes the covariant derivative defined with the independent
connection $ \Gamma^{\lambda}_{\mu\nu} $, and $ (\mu\nu) $ denote symmetrization over the indices $ \mu $ and $ \nu $.

Next, we examine the role of $ {\Delta_{\lambda}}^{\mu\nu} $ . By
splitting eq. (8) into a symmetric and an antisymmetric part and
performing contractions and manipulations it can be shown that
\cite{c10} \b {\Delta_{\lambda}}^{[\mu\nu]}=0 \Rightarrow
{S_{\mu\nu}}^{\lambda}=0.\e This straightforwardly implies two
things: a) Any torsion is introduced by matter fields for which $
{\Delta_{\lambda}}^{[\mu\nu]} $ is non-vanishing; b) torsion is
not propagating, since it is given algebraically in terms of the
matter fields through $ {\Delta_{\lambda}}^{[\mu\nu]} $. It can,
therefore, only be detected in the presence of such matter fields.
In the absence of the latter, space-time will have no torsion.

Obviously, there are certain types of matter fields for which $
{\Delta_{\lambda}}^{\mu\nu} $ = 0. Characteristic example is: A
scalar field, since in this case the covariant derivative can be
replaced with a partial derivative. Therefore, the connection does
not enter the matter action. On the contrary, particles with spin,
such as Dirac fields, generically have a non-vanishing
hyper-momentum and will, therefore, introduce torsion. A more
complicated case is that of a perfect fluid with vanishing
vorticity. If we set torsion aside, or if we consider a fluid
describing particles that would initially not introduce any
torsion then, as for a usual perfect fluid in GR, the matter
action can be written in terms of three scalars: the energy
density, the pressure, and the velocity potential\cite{c16}.
Therefore such a fluid will lead to a vanishing $
{\Delta_{\lambda}}^{\mu\nu} $ . However, complications arise when
torsion is taken into account: Even though it can be argued that
the spins of the individual particles composing the fluids will be
randomly oriented, and therefore the expectation value for the
spin should add up to zero, fluctuations around this value will
affect space-time\cite{c10}. Of course, such effects will be
largely suppressed, especially in situations in which the energy
density is small, such as late-time cosmology.

\section{Modified expression of Mach's principle }

Curvature in the Einstein general relativity is one of the main
concepts that to be considered and all calculations, related to
the field equations, return to curvature. Curvature can explain
experimental observations such as the motion of planets in the
solar system, time delay, bending of light near the stars, and the
convergence of light in lensing effect. Therefore, Mach's
principle in general relativity, based on the concept of
curvature, expressed as follows: \textbf{\textit{''Matter tells
space-time how to curve.''}}

On the other hand, in the previous section, it shown that
metric-affine $ f(R) $ gravity allows the presence of torsion.
Torsion is merely introduced by specific forms of matter; those
for which the matter action has a dependence on the connections.
Therefore, the form of Mach's principle is corrected as follows:
as ''matter tells space-time how to curve'',
\textbf{\textit{''matter will also tell space-time how to
twirl''}} \cite{c10}. But we also do not accept this new sentence,
because we believe that torsion has the more comprehensive concept
than twirl and curvature.

For a more clear this issue, we consider how to move a bolt while
being wound on a wooden surface. If a bit of pressure put on it,
just the tip of the bolt rotates (Two-dimensional motion) on a
wooden surface and in this case it just means twirl. But if we put
more pressure on it, along with rotation of bolt tip, it can also
penetrate into the wooden surface (Third dimension of the
transition can move perpendicular to the surface or not).
Obviously, with increasing the pressure on the screw bolt, the
bolt goes into the wood and in this case we see only a hole on the
wooden surface. In this example, the rotary motion to add a
transitional move can be represent torsion is more accurate. Note
that the concepts of this example can be generalized to the
super-surface and space-time with the higher dimensions.

According to the above physical interpretation for torsion, it
seems that the high energy area-near black holes and at the Planck
energy limit- torsion of space-time is more realistic than twirl
and curvature. For example, the intense of gravitational
attraction around the black hole horizon, first, to divert the
direction of objects and light (bending in the path). Then due to
the increased gravitational force for objects closer to the
horizon, they are rotating around the center of the black hole.
Finally, when objects are passed through the horizon, fall into
the black hole and are swallowed. Consequently, due to the
appearance of torsion of space-time in the high energy range, we
can also modify expression of Mach's principle in reference
\cite{c10} as follows:\textbf{\textit{''Matter will also tell
space-time how to twist".}} We also recommend that the
metric-affine formalism is more likely to be introduced in the
high energy physics regions.

\section{Discussions and Conclusions}

In this paper, we first overview the formalism of $ f(R) $
gravity. Then we take a closer look in the metric-affine formalism
and the concept of torsion of space-time. If one accepts the
interpretation presented in this paper for torsion of space-time,
then as it was discussed, torsion will play a major role in
formation of the geometry of space-time near the black holes and
the early universe cosmology. Thus the metric-affine formalism is
more likely to be introduced in these regions and when consider
the lower energy limit, we can make use of other formalisms, i.e.
the metric and Palatiny formalisms. It seems that the study of
metric-affine formalism and torsion of space-time in the high
energy physics require a lot of research in the future.

\noindent{\bf{Acknowlegements}}: This work has been supported by the
Islamic Azad University-Qom Branch, Qom, Iran.

\end{document}